\documentclass[prd,aps,tightline,twocolumn,nofootinbib,superscriptaddress]{revtex4}
\usepackage{graphicx}
\usepackage{amssymb}
\usepackage{amsmath}

\newcommand{\bear}{\begin{array}}  
\newcommand {\eear}{\end{array}}
\newcommand{\bea}{\begin{eqnarray}}   
\newcommand{\eea}{\end{eqnarray}}
\newcommand{\beq}{\begin{equation}}   
\newcommand{\eeq}{\end{equation}}
\newcommand{\bef}{\begin{figure}}  \newcommand 
{\eef}{\end{figure}}
\newcommand{\bec}{\begin{center}}  \newcommand 
{\eec}{\end{center}}

\begin{document}

%\begin{titlepage}

%\begin{flushright}
%IPMU 08-0112 \\
%ICRR-Report-535
%\end{flushright}

\title{
Neutrino Signals from Annihilating/Decaying Dark Matter \\in the Light of Recent 
Measurements of Cosmic Ray Electron/Positron Fluxes
}

\author{Junji Hisano} 
\affiliation{Institute for Cosmic Ray Research, University of Tokyo, Kashiwa 277-8582, Japan}
\affiliation{Institute for the Physics and Mathematics of the Universe,
University of Tokyo, Kashiwa 277-8568, Japan}

\author{Masahiro Kawasaki}
\affiliation{Institute for Cosmic Ray Research, University of Tokyo, Kashiwa 277-8582, Japan}
\affiliation{Institute for the Physics and Mathematics of the Universe, University of Tokyo, Kashiwa 277-8568, Japan}

\author{Kazunori Kohri}
\affiliation{Physics Department, Lancaster University, Lancaster LA1 4YB, UK}

\author{Kazunori Nakayama}
\affiliation{Institute for Cosmic Ray Research, University of Tokyo, Kashiwa 277-8582, Japan}

\date{\today}

\begin{abstract} The excess of cosmic-ray electron and positron fluxes
  measured by the PAMELA satellite and ATIC balloon experiments may be
  interpreted as the signals of the dark matter annihilation or decay
   into leptons.  In this letter we show that the dark matter
  annihilation/decay which reproduces the electron/positron excess may
  yield a significant amount of high-energy neutrinos from the Galactic center. In the
  case, future kilometer-square size experiments may confirm such a
  scenario, or even the Super-Kamiokande results already put
  constraints on some dark matter models.
\end{abstract}

%\preprint{IPMU-08-???}
%\preprint{ICRR-Report-???}
%\pacs{98.80.Cq}

\maketitle

%\end{titlepage}

Recently the PAMELA satellite experiment reported a clear excess of
the positron flux from the expected background \cite{Adriani:2008zr},
and the ATIC \cite{:2008zz} and PPB-BETS \cite{Torii:2008xu} balloon
experiments have shown the rise in the total electron and positron
flux.  The HESS collaboration also released a data of the electron
flux \cite{Collaboration:2008aa}, which is consistent with
ATIC/PPB-BETS results.  While those results may be explained by the
contribution from the pulsar(s) \cite{Atoian:1995ux,Hooper:2008kg}, they may be
interpreted as the signatures of the annihilation/decay of the dark
matter. Many papers have appeared on the later subject
\cite{Bergstrom:2008gr,Bertone:2008xr}.  

If the observed electron/positron excesses come from the dark matter
annihilation, a large annihilation cross section of order of $\sim
10^{-24}$-$10^{-23}~{\rm cm^3s^{-1}}$ is required, slightly depending
on the annihilation mode.  Otherwise, a huge boost factor must be
introduced, especially when we stick to the annihilation cross section
of $\sim 3\times 10^{-26}~{\rm cm^3s^{-1}}$, which accounts for the
present dark matter abundance under the standard thermal relic
scenario \cite{Jungman:1995df}. The annihilation/decay of the dark
matter also yields gamma rays, anti-protons and synchrotron radiations
in the Galaxy. Some models proposed to explain the electron/positron
excesses may be disfavored due to the saturation of those observed
limits. It was also shown that the dark matter annihilation scenario
as an explanation of the positron excess is constrained by the
big-bang nucleosynthesis \cite{Hisano:2008ti}.  
 Thus, it is important to look for observational
signatures of the annihilation/decay of the dark matter which may be
related to the electron/positron fluxes in order to distinguish or
exclude some particular scenarios.

In this letter, we evaluate the neutrino flux from the Galactic
center, which come from the dark matter annihilation/decay into
  leptons.  Such dark matters are 
favorable since excess is not observed in the anti-proton cosmic rays
spectrum. we point out that observations of high-energy
neutrinos arising from the dark matter annihilation/decay into leptons
at the Galactic center can provide constraints 
even at the present stage, and future neutrino telescope projects in
the northern hemisphere, such as KM3NeT, will be useful for a cross
check of the dark matter scenario, if the currently observed
electron/positron excesses truly originate from the dark matter. In
particular, we show that the study of neutrino-induced up-going muon flux is
very important for a heavy dark matter with mass of order of a few TeV
indicated by ATIC/PPB-BETS results, as opposed to the case of rather
light dark matter, which was studied in Ref.~\cite{Covi:2008jy} for
the specific model of decaying gravitino dark matter with $R$-parity
violation.

%%%%%%%%%%%%%%%%%%%%%%%%%%%%%%%

First, we briefly review the electron and positron flux produced by
the dark matter annihilation/decay \cite{Baltz:1998xv}.  Since
high-energy electrons and positrons produced by the dark matter
annihilation/decay lose their energy quickly due to the synchrotron
emission induced by the Galactic magnetic field and inverse Compton
processes with CMB photons and star light, they can reach to the Earth
only from the region within a few kpc.  The propagation of electrons
and positrons is described by the following diffusion equation,
\begin{equation}
\begin{split}
        \frac{\partial}{\partial t}f(E, \vec x) 
        = &K(E)\nabla^2f(E, \vec x) \\
        &+\frac{\partial}{\partial E} [b(E)f(E, \vec x)] + Q(E,\vec x), \label{diffusion}
\end{split}
\end{equation}
where $f(E, \vec x)$ denotes the electron and positron number density
at $\vec x$ with energy $E$. The flux at the Earth (${\vec x}={\vec
  x_\odot}$) is given by $\Phi^{(\rm DM)}_{e^-,e^+}(E,\vec
x_\odot)=(c/4\pi)f(E,\vec x_\odot)$ with the speed of light $c$.
$K(E)$ and $b(E)$ represent the diffusion constant and energy-loss
rate, respectively.  The source term $Q(E,\vec x)$ is given by
\begin{equation}
        Q(E,\vec x) = \frac{1}{2} \frac{\rho^2(\vec x)}{m_{\chi}^2} \sum_F \langle \sigma v \rangle_F 
        \frac{dN^{(e^-,e^+)}_F}{dE},
\end{equation}
for the case of annihilation, and
\begin{equation}
        Q(E,\vec x) = \frac{\rho(\vec x)}{m_{\chi}} \sum_F \Gamma_F 
        \frac{dN^{(e^-,e^+)}_F}{dE},
\end{equation}
for the case of decay. Here, $\rho(\vec x)$ is the mass density of the
dark matter, $m_\chi$ is the dark matter particle mass, $\langle
\sigma v \rangle_F$ and $\Gamma_F$ are the annihilation cross section
and the decay rate into the final state $F$ respectively, {\bf and
$dN^{(e^-,e^+)}_F/dE$} is the fragmentation function of the final state
$F$ into electrons/positrons.  The steady-state solution of this
equation can be obtained semi-analytically with a cylinder-like
boundary condition \cite{Hisano:2005ec}.

\begin{figure}[t]
 \begin{center}
   \includegraphics[width=1.0\linewidth]{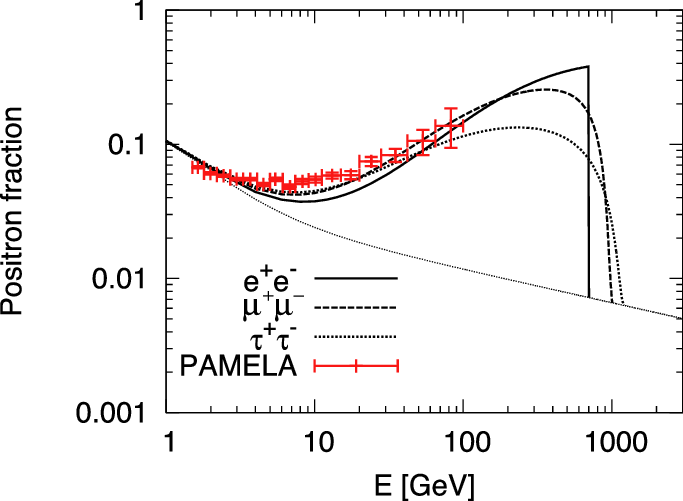} 
   \includegraphics[width=1.0\linewidth]{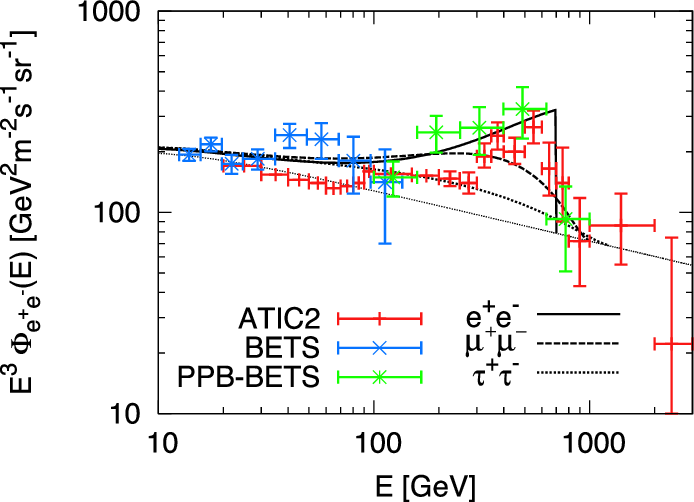} 
   \caption{ The positron fraction (top) and
   total electron and positron flux from the annihilation (bottom)
   from dark matter annihilation.
   We take the mass and annihilation cross section as $m=0.7~$TeV and
   $\langle \sigma v\rangle =5\times 10^{-24}~{\rm cm^3s^{-1}}$ for the mode into 
   $e^+ e^-$ (solid),
   $m=1~$TeV and $\langle \sigma v\rangle =1.5\times 10^{-23}~{\rm cm^3s^{-1}}$ 
   for the mode into $\mu^+ \mu^-$ (dashed),
   $m=1.2~$TeV and $\langle \sigma v\rangle =2\times 10^{-23}~{\rm cm^3s^{-1}}$ 
   for the mode into $\tau^+ \tau^-$ (dotted).
   Results of PAMELA, ATIC2, BETS and PPB-BETS are plotted.}
   \label{fig:eflux}
 \end{center}
\end{figure}

Fig.~\ref{fig:eflux} shows positron fraction (top) and total electron
and positron flux from the dark matter annihilation (bottom).  We take
the mass and annihilation cross section as $m=0.7~$TeV and $\langle
\sigma v\rangle =5\times 10^{-24}~{\rm cm^3s^{-1}}$ for the mode into
$e^+ e^-$ (solid), $m=1~$TeV and $\langle \sigma v\rangle =1.5\times
10^{-23}~{\rm cm^3s^{-1}}$ for the mode into $\mu^+ \mu^-$ (dashed),
$m=1.2~$TeV and $\langle \sigma v\rangle =2\times 10^{-23}~{\rm
  cm^3s^{-1}}$ for the mode into $\tau^+ \tau^-$ (dotted).  We have
added a background flux given in Ref.~\cite{Baltz:1998xv} with
normalization fixed by hand.  Also plotted are results of PAMELA
\cite{Adriani:2008zr}, ATIC \cite{:2008zz}, BETS \cite{Torii:2001aw}
and PPB-BETS \cite{Torii:2008xu}.  We can see that these models well
fit the observed data.  Typical cross section into leptons which
reproduces the observational data is around $\langle \sigma v\rangle
\sim10^{-23}~{\rm cm^3s^{-1}}$, which may be the result of Sommerfeld
enhancement \cite{Hisano:2003ec}.  Quite similar results are obtained
for the case of decaying dark matter.  In that case, the typical decay
rate into leptons should be $\sim 10^{-26}~{\rm s^{-1}}$ with the mass
around 1.5-3.0~TeV.

%%%%%%%%%%%%%%%%%%FIGURE%%%%%%%%%%%%%%%%%%%

%%%%%%%%%%%%%%%%%%%%%%%%%%%%%%%%%%%%%%%%%

Now let us discuss the production of neutrinos from the
annihilation/decay of the dark matter particle.  As is already mentioned,
dark matter annihilation/decay directly into leptons are more favored,
and the typical annihilation cross section and decay rate required for
explaining the electron/positron excesses are $\langle \sigma v\rangle
\sim10^{-23}~{\rm cm^3s^{-1}}$ for the case of annihilation, and
$\Gamma \sim 10^{-26}~{\rm s}^{-1}$ for the case of decay.  When
  the dark matter is $SU(2)\times U(1)$ singlet and annihilates/decays
  into left-handed charged leptons, the dark matter also
  annihilates/decays into neutrinos with same cross section/decay rate.
  Furthermore, charged leptons
  ($\mu, \tau$) produced in the annihilation/decay of the dark matter
  decay into neutrinos.  Thus it is natural to expect that when the
dark matter mainly annihilates or decays into leptons, it also
produces comparable amount of neutrinos.  Interestingly, such a value
of the cross section is close to the upper bound obtained from the
neutrino flux assuming that the dark matter totally annihilates into
neutrinos \cite{Beacom:2006tt}.

The possible production processes of neutrinos are direct production
($\chi (+\chi) \to \nu_i +\bar \nu_i$) where $i=1,2,3$ distinguishes
flavors and the decay of primary $\mu$'s and $\tau$'s ($\mu^- \to
\nu_\mu + \bar \nu_e + e^-$, etc.)  which are directly produced by the
dark matter annihilation/decay ($\chi (+\chi) \to \mu^-+\mu^+, \tau^-
+\tau^+$).  The neutrino flux at the Earth coming from the Galactic
center is evaluated by
\begin{equation}
	\frac{dF_{\nu_i}}{dE}=\frac{R_\odot \rho_\odot^2}{8\pi m_\chi^2}
		\left (\sum_F \langle \sigma v\rangle _F \frac{dN_F^{(\nu_i)}}{dE} \right )
		\langle J_2 \rangle_\Omega \Delta \Omega,
\end{equation}
for the case of annihilation, and
\begin{equation}
	\frac{dF_{\nu_i}}{dE}=\frac{R_\odot \rho_\odot}{4\pi m_\chi}
		\left (\sum_F \Gamma _F \frac{dN_F^{(\nu_i)}}{dE} \right )
		\langle J_1 \rangle_\Omega \Delta \Omega,
\end{equation}
for the case of decay. Here, $R_\odot = 8.5$~kpc and
$\rho_\odot=0.3~{\rm GeV cm^{-3}}$ are the distance of the solar
system from the Galactic center and local dark matter density near the
solar system, $F$ collectively denotes the primary annihilation/decay mode
(e.g., $\mu^+ \mu^-$, etc.), and $dN_F^{(\nu_i)}/dE$ represents the
neutrino spectrum arising from the final state $F$.  The dependence on
the dark matter halo density profile is contained in the remaining
factor $\langle J_n \rangle_\Omega$, defined by
\begin{equation}
	\langle J_n \rangle_\Omega = \int \frac{d\Omega}{\Delta \Omega}
	\int_{\rm l.o.s.}\frac{dl(\psi)}{R_\odot}\left ( \frac{\rho(l)}{\rho_\odot} \right )^n,
\end{equation}
where $l(\psi)$ is the distance from us along the direction $\psi$,
which is the cone-half angle from the Galactic center within the range $0<\psi<\psi_{\rm max}$, and 
$\Delta \Omega(\equiv 2\pi (1-\cos\psi_{\rm max}))$ 
is the solid angle over which the neutrino flux is averaged.
Typical values are $\langle J_2 \rangle_\Omega \Delta \Omega \sim 10$ and
$\langle J_1 \rangle_\Omega \Delta \Omega \sim 0.4$ for $\psi_{\rm max} =5^\circ$,
if the Navarro-Frenk-White (NFW) halo density profile \cite{Navarro:1995iw} is adopted.  
For the isothermal profile
$\langle J_{2(1)} \rangle_\Omega \Delta \Omega$ is $\sim 2(0.4)$.

The best technique for detecting neutrinos from the Galactic center is
observation of the up-going muons, which are produced in the rock
below the detectors. The detectors in the northern hemisphere observe
the up-going muon induced by neutrinos from the Galactic center.  The
neutrino-induced muon flux is evaluated from the neutrino flux
\cite{Ritz:1987mh,Jungman:1995df} as
\begin{equation}
\begin{split}
	F_{\mu^+ \mu^-}^{(\rm ann)} &\simeq 5.9\times 10^{-15}~{\rm cm^{-2}s^{-1}} \\
	&\times \sum_F S_F
	\left ( \frac{\langle \sigma v\rangle _F }{10^{-23}~{\rm cm^3s^{-1}}} \right )
	\left ( \frac{\langle J_2 \rangle_\Omega \Delta \Omega}{10} \right ),   \
\label{muann}
\end{split}
\end{equation}
for the case of annihilation and
\begin{equation}
\begin{split}
	F_{\mu^+ \mu^-}^{(\rm dec)} &\simeq 2.0\times 10^{-15}~{\rm cm^{-2}s^{-1}} 
	\left ( \frac{m_\chi}{2~{\rm TeV}} \right )\\
	&\times \sum_F S_F
	\left ( \frac{\Gamma _F }{10^{-26}~{\rm s^{-1}}} \right )
	\langle J_1 \rangle_\Omega \Delta \Omega,   \label{mudec}
\end{split}
\end{equation}
for the case of decay.  Here we have defined $S_F$ through
\begin{equation}
	S_F = \sum_{\nu_i}\int_{E_{\rm min}}^{E_{\rm in}}\frac{dN_F^{(\nu_i)}}{dE}P_{\nu_i \nu_\mu} 
	\left ( \frac{E}{E_{\rm in}} \right )^2 dE,
\end{equation}
where $E_{\rm in}=m_\chi(m_\chi/2)$
for the case of annihilation (decay), and
$E_{\rm min}$ is the threshold energy above which the muons can be
detected. $P_{\nu_i\nu_\mu}$ denotes the probability that the
$\nu_i$ at the production is observed as $\nu_\mu$ at the Earth due to
the effect of neutrino oscillation.  The value of $S_F$ is summarized
in a following table.
These values are insensitive to the precise value of $E_{\rm min}$
as long as it is smaller than 100~GeV.
\begin{center}
\begin{tabular}{|c||c|c|c|c|c|c|c|} 
\hline
{\rm } 
& $\nu_e\bar{\nu}_e$
& $\nu_\mu \bar{\nu}_\mu $
& $\nu_\tau \bar{\nu}_\tau $
& $\mu_R^-\mu_L^+$
& $\mu_L^-\mu_R^+$
& $\tau_R^-\tau_L^+$
& $\tau_L^-\tau_R^+$
\\
\hline
$S_F$
& 0.44
& 0.78
& 0.78
& 0.19
& 0.20
& 0.14
& 0.18
\\
\hline
\end{tabular}
\end{center}

Remarkably, the neutrino-induced muon flux is proportional to the
second moment of the neutrino energy spectrum. It implies that the
up-going muon detection is more sensitive to the energetic
neutrinos. As the result, the muon flux is independent of the dark
matter mass in the case of annihilation, and it is proportional to the
mass in the case of decay. (See Eqs.~(\ref{muann},\ref{mudec}).)
This is a good news because ATIC/PPB-BETS indicate TeV
scale dark matter mass, rather than of the order of 100~GeV, and such
heavy dark matter models have benefits from the viewpoint of detection
at the neutrino detectors.

%%%%%%%%%%%%%%%%%FIGURE%%%%%%%%%%%%%%%%%%%

\begin{figure}[t]
 \begin{center}
   \includegraphics[width=1.0\linewidth]{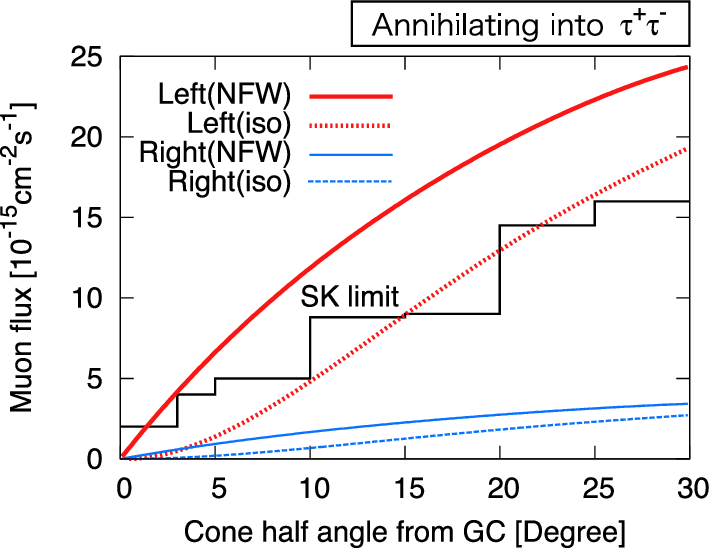}
   \includegraphics[width=1.0\linewidth]{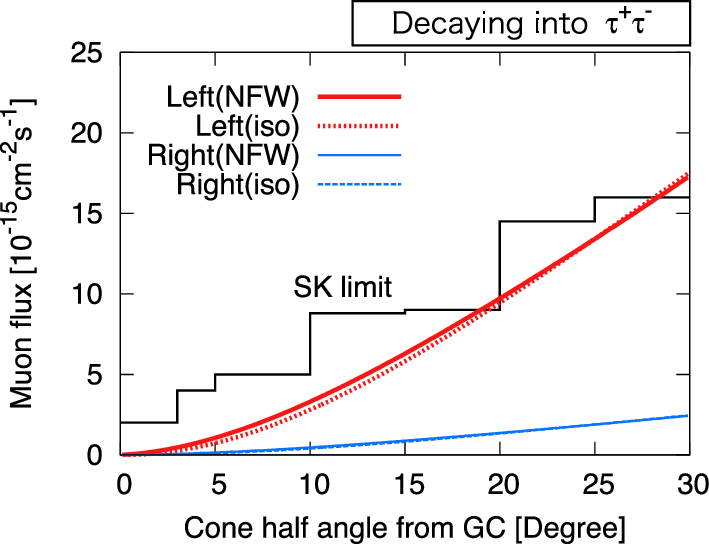} 
   \caption{Expected up-going muon flux from the dark matter
     annihilation (top) and decay (bottom) as a function of the cone
     half angle from the Galactic center.  SK limits are also shown.
     Here, we assume the the dark matter annihilates/decays into
     left-handed leptons ($\tau_L^- \tau_R^+$ and $\nu_\tau \bar
     \nu_\tau$) shown by ``Left'' or right-handed leptons ($\tau_R^-
     \tau_L^+$) shown by ``Right''.  Annihilating Dark matter models
     correspond to those used in Fig.~\ref{fig:eflux}. For the case of
     decaying dark matter, we used $\Gamma = 1.5\times 10^{-26}~{\rm
       s}^{-1}$ and $m=2.4~$TeV.  The dark matter density profile is
     assumed to be the NFW and isothermal profiles.  }
   \label{fig:tau}
 \end{center}
\end{figure}

%%%%%%%%%%%%%%%%%%%%%%%%%%%%%%%%%%%%%%%%%

%%%%%%%%%%%%%%%%%FIGURE%%%%%%%%%%%%%%%%%%%

\begin{figure}[t]
 \begin{center}
   \includegraphics[width=1.0\linewidth]{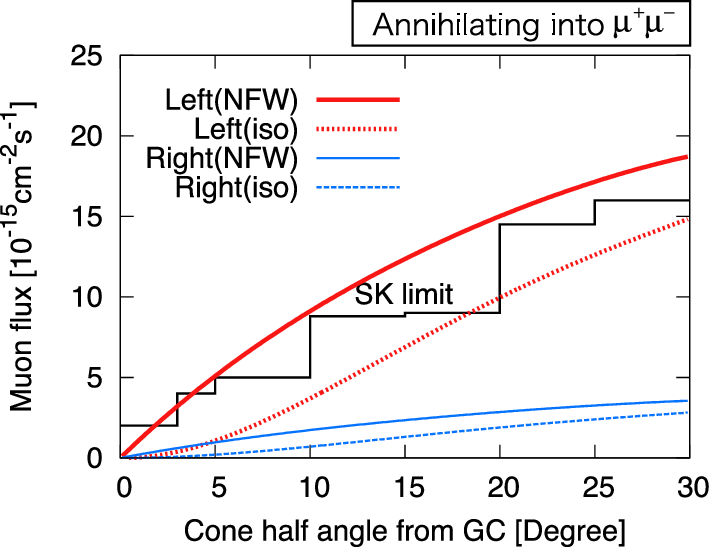} \\
   \includegraphics[width=1.0\linewidth]{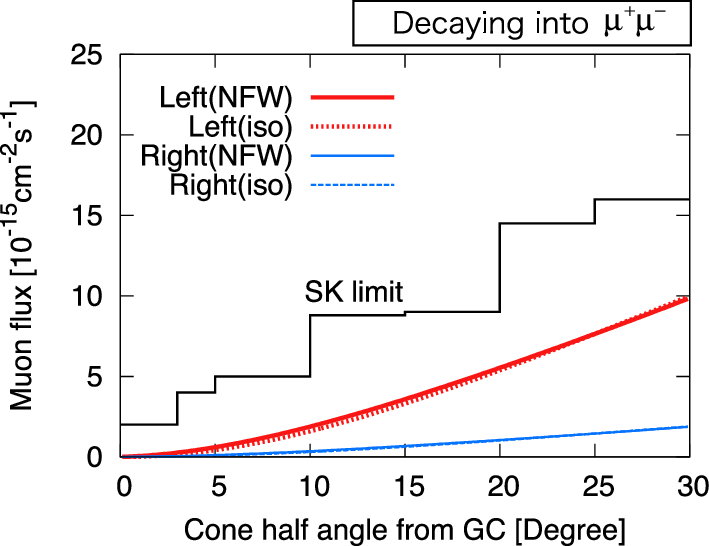} 
   \caption{ Same as Fig.~\ref{fig:tau}, but for annihilating (decaying) into $\mu^- \mu^+$.
For the case of
     decaying dark matter, we used $\Gamma = 1\times 10^{-26}~{\rm
       s}^{-1}$ and $m=2~$TeV.  
	}
   \label{fig:mu}
 \end{center}
\end{figure}

%%%%%%%%%%%%%%%%%%%%%%%%%%%%%%%%%%%%%%%%%

%%%%%%%%%%%%%%%%%FIGURE%%%%%%%%%%%%%%%%%%%

\begin{figure}[t]
 \begin{center}
   \includegraphics[width=1.0\linewidth]{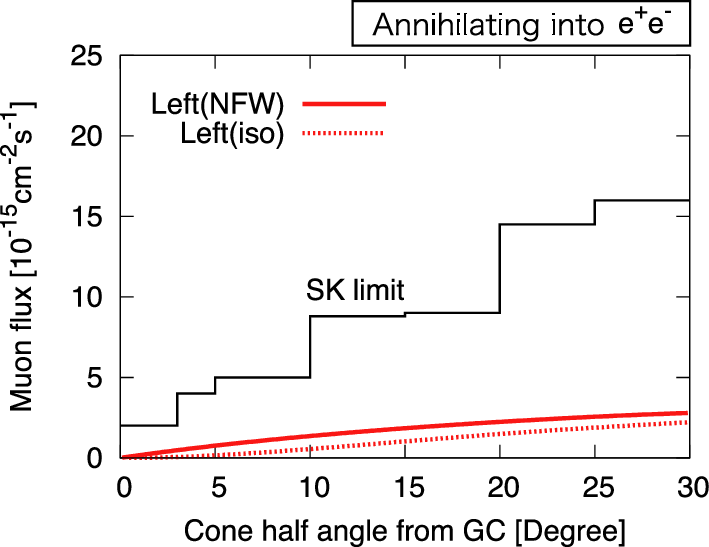} 
   \includegraphics[width=1.0\linewidth]{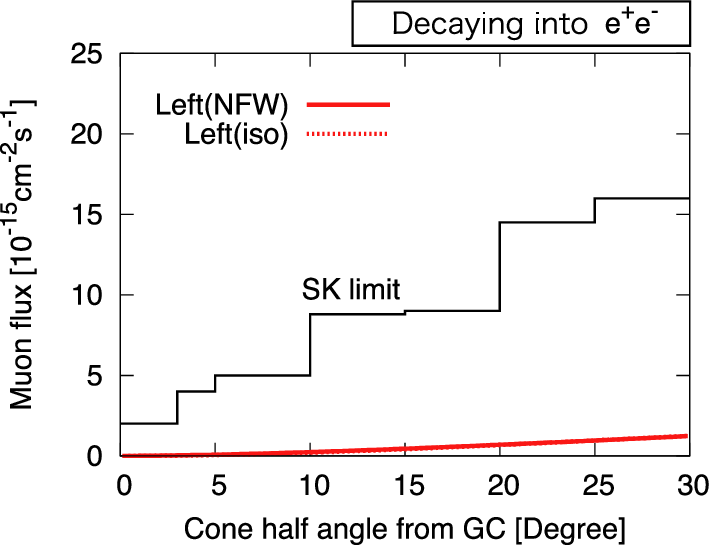} 
   \caption{ Same as Fig.~\ref{fig:tau}, but for annihilating (decaying) into $e^-e^+$. For the case of
     decaying dark matter, we used $\Gamma = 4\times 10^{-26}~{\rm
       s}^{-1}$ and $m=1.4~$TeV.  
   }
   \label{fig:e}
 \end{center}
\end{figure}

%%%%%%%%%%%%%%%%%%%%%%%%%%%%%%%%%%%%%%%%%

In order to reproduce ATIC/PPB-BETS anomaly, dark matter
annihilation/decay directly into leptons are more favored as
  mentioned above.
Here, we assume for simplicity that 
the dark matter particle is $SU(2)\times U(1)$ singlet. In the case,
monochromatic neutrinos are naturally generated
when the dark matter annihilates/decays into left-handed charged
leptons. On the other hand, the dark matter 
annihilating/decaying into only the right-handed charged leptons can 
produce neutrinos secondarily.
 Therefore, in the following, we consider two cases: dark
matter annihilates (decays) into left-handed leptons (i.e., neutrinos
and charged leptons) with same branching ratio, and into all
right-handed-leptons (i.e., no direct production of neutrinos).

The muon flux from the dark matter annihilation (decay) is shown in
the top (bottom) panel of Figs.~\ref{fig:tau}-\ref{fig:e}, as a
function of the cone half angle from the Galactic center, both for the
isothermal and NFW profile and for the case of annihilation (decay)
into $\tau^+\tau^-$, $\mu^+\mu^-$ and $e^+e^-$.  We have assumed same
mass and cross section as those used in Fig.~\ref{fig:eflux} so that
produced charged leptons exhibit a good fit on the PAMELA and
ATIC/PPB-BETS results.  For the case of decaying dark matter, we used
following parameters : $\Gamma = 4\times 10^{-27}~{\rm s}^{-1}$ and
$m=1.4~$TeV for the case of $e^+e^-$, $\Gamma = 1\times 10^{-26}~{\rm
  s}^{-1}$ and $m=2~$TeV for the case of $\mu^+\mu^-$, $\Gamma =
1.5\times 10^{-26}~{\rm s}^{-1}$ and $m=2.4~$TeV for the case of
$\tau^+\tau^-$.  Resulting positron and electron flux looks quite
similar to Fig.~\ref{fig:eflux}.  In the
case of annihilation (decay) into left-handed leptons, we have assumed
the same annihilation (decay) rate into neutrino pairs.  Also shown
are limits from Super-Kamiokande (SK) \cite{Desai:2004pq}.  It is
found that if the neutrinos are directly produced by the
annihilation/decay of the dark matter with the same rate as charged
leptons, SK may already give constraints on some dark matter models
such as those annihilating/decaying into $\tau ^+ \tau^-$ and 
annihilating into $\mu^+ \mu^-$.  Notice
that parameters chosen here (annihilation cross section, decay rate
and mass) should be regarded as only representative values, and they
include $\mathcal O(1)$ uncertainties in order to fit the PAMELA/ATIC
results depending on dark matter models, which may either increase or
decrease the resulting muon flux.  Therefore, model-by-model
comparison should be performed in order to check a consistency with SK
bound.

Some comments are in order.  The final state leptons radiate
gamma rays through the internal bremstrahlung or cascade decay
processes, which should be compared with the gamma-ray flux observed
by the HESS experiment \cite{Aharonian:2004wa}.  In the case of
annihilation, the constraint is severe and the PAMELA/ATIC results may
be inconsistent with the HESS observation, if the cuspy density
profile such as NFW profile is adopted \cite{Bertone:2008xr}.  This
can be relaxed for the isothermal profile.  On the other hand, the
dependence of the neutrino flux on the density profile is rather weak
since the SK looks at the Galactic center over wide angle.  Thus it
may be possible that neutrinos give more stringent constraint on the
dark matter annihilation scenario.  If only the PAMELA anomaly is
taken into account, lighter dark matter is possible.  In that case,
the muon signals at neutrino detectors are suppressed since low-energy
neutrinos have less potential to be converted into muons and less
propagation length inside the Earth.

In the case of decaying dark matter, neutrino constraints are very
useful since the gamma-ray flux gives only loose constraints even for
the NFW profile \cite{Bertone:2008xr}.  In addition, neutrino-induced
muon flux receives an extra enhancement factor proportional to the
dark matter mass as shown in Eq.~(\ref{mudec}).  Therefore, a decaying
dark matter with a few TeV, indicated by ATIC results, may have
distinct signatures on the neutrino flux, rather than gamma rays.

It is noticed that a significant amount of neutrino flux is not
expected in the pulsar scenario for the positron excess.  Therefore,
if we detect unknown neutrino signals from the Galactic center, the
annihilation/decay scenario could be distinguished from the other
astrophysical scenarios and be verified.  The planned future mega-ton
scale water tank detector, Hyper-Kamiokande, which is the extension of
the SK, and kilo-meter size detector, such as KM3NeT, are expected to
improve the current sensitivity of the SK by one or two orders of
magnitude.  It is recently discussed that the IceCube DeepCore, which
is a planned extension of the IceCube, would also have sensitivity on
high-energy neutrinos from the Galactic center \cite{Cowen:2008zz}.
These neutrino experiments will be very useful for confirming,
distinguishing or excluding some of the dark matter models as an
explanation of the recently observed cosmic electron/positron
excesses.

%%%%%%%%%%%%%%%%%%%%%%%%%%%%%%%%%%%%%%%%%%%%%%%
\begin{acknowledgements}
%%%%%%%%%%%%%%%%%%%%%%%%%%%%%%%%%%%%%%%%%%%%%%%

The authors appreciate Prof.~Kajita who encouraged them to evaluate the
  constraints on neutrino flux originated from the dark matter annihilation and decay. 
 KN would like to thank the Japan Society
  for the Promotion of Science for financial support.  This work is
  supported by Grant-in-Aid for Scientific research from the Ministry
  of Education, Science, Sports, and Culture (MEXT), Japan,
  No.14102004 (MK) and No.~20244037 and No.~2054252 (JH), and also by
  World Premier International Research Center InitiativeiWPI
  Initiative), MEXT, Japan.  KK is supported in part by PPARC grant,
  PP/D000394/1, EU grant MRTN-CT-2006-035863, the European Union
  through the Marie Curie Research and Training Network
  ``UniverseNet''.

%%%%%%%%%%%%%%%%%%%%%%%%%%%%%%%%%%%%%%%%%%%%%%%
\end{acknowledgements}
%%%%%%%%%%%%%%%%%%%%%%%%%%%%%%%%%%%%%%%%%%%%%%%

%%%%%%%%%%%%%%%%%%%%%%%%%%%%%%%%%%%%
%\section*{Appendix} \label{app}
%%%%%%%%%%%%%%%%%%%%%%%%%%%%%%%%%%%%

%\appendix

%%%%%%%%%%%%%%%%%%%%%%%%%%%%%%%%%%%%
{}
%%%%%%%%%%%%%%%%%%%%%%%%%%%%%%%%%%%%

\end{document}